\documentclass[%
 reprint,
 amsmath,amssymb,
 aps,
]{revtex4-2}
\usepackage{graphicx}
\usepackage{bm}
\usepackage{hyperref}

\usepackage{amsmath}
\usepackage{slashed}

\begin{document}


\title{The Majorana Hopping and Constraints in the Anti-de Sitter Spacetime}
\author{Chia-Li Hsieh}%
 \email{galise@gmail.com}

\author{Vahideh Memari}
 \altaffiliation[]{vahideh.memari@emu.edu.tr}

\author{Mustafa Halilsoy}
 \altaffiliation[]{mustafa.halilsoy@emu.edu.tr}

\affiliation{%
 Department of Physics, Faculty of Arts and Sciences, Eastern Mediterranean University, Famagusta,
North Cyprus via Mersin 10, Turkey 
}%


\date{\today}

\begin{abstract}
Comformally flat spacetime converts the Anti-de Sitter (AdS) spacetime into a z-dependent metric. We take advantage of this metric to solve Dirac equations in the AdS geometry analytically, including Dirac, Majorana, Dirac tachyon and Majorana tachyon. Constraints on Majorana wavefunctions limit the mass. Besides, there gives rise flatlands and Majorana particles hop in between, just mimicking the D-brane stack and open strings. Furthermore, a plausible way of ion-trapped experiment is suggested.

 .

\end{abstract}

\keywords{Suggested keywords}
\maketitle



\section{\label{sec:level1}Introduction}

The Dirac equation in Anti-de Sitter(AdS), de Sitter \cite{Dirac1935} or other geometries (please see Chandrasekhar's book \cite{Chandrasekhar1983}) is not an old issue.
The task to find solutions of Dirac equations relies on decoupling the equations, especially in curved spacetime.
Instead of working in spherical coordinate \cite{Bachelot2008}, in this letter, however we notice that conformally flat transformation of the AdS spacetime simplifies the discussion greatly.
The metric and Dirac equation become z-dependent only.
Then with the help of Newman-Penrose (NP) formalism \cite{Newman1962} and tricks of decoupling in section II, we decouple the equations neatly.

Theoretically, we argue that the Dirac particles possess superluminal behavior and self-complex conjugate, the Majorana condition. Exceeding the speed of light with imaginary mass, as Chodos has proposed in neutrino physics \cite{Chodos2023}, seems unnatural but some experiments with effective $im$ mass is plausible. Ion trap is one of them to enable to reach the quantum relativistic effects. Teams with Solano \cite{Gerritsma2010} have illustrated the simulation of Dirac particles and the Zitterbewegung, trembling motions of Dirac fermions. By introducing a non-Hermitian term in the Hamiltonian, the ion-trapped Dirac tachyon is also studied \cite{Lee2015}. They move fast than the effective speed of light with imaginary mass. The motivation of studying tachyonic fermions comes from this effective $im$ mass. In this paper, We shall not only impose the Dirac tachyon \cite{Lemke1976} but also consider Majorana tachyon in the AdS geometry. Peculiarly, we found a 2+1-D flatland embedded in the curved 3+1-D AdS spacetime, which is not yet seen in the AdS metric. Interestingly, particles can hop in between the flatlands, just like the relation between D-brane stack and open strings \cite{Dai1989}. Furthermore, in some conditions, the mass of Majorana particles is constrained. We then go further to think about a feasible method to mimic AdS geometry with trapped ions as fermions in the lab.

Our paper is organized as follows. In section II, we discuss the differential equations and derive the wavefunctions and the boundary. We consider the tachyonic fields in the AdS background in section III. In Section IV, the Majorana condition is imposed. We complete the paper with our discussion and conclusion in section V.

\section{The Method}

Physically, AdS spacetime is a solution of Einstein’s field equation with a negative constant $\Lambda$, giving an attracting geometry. And the metric can be read as
$ds^2=(1+\frac{\Lambda}{3}r^2)d\bar{t}^2-\frac{dr^2}{1+\frac{\Lambda}{3}r^2}-r^2(d\theta^2+\sin^2\theta d\phi^2)$, with cosmological constant $\Lambda$. We can transform it into a conformally flat spacetime without changing physical meaning (just seen from another frame)
\begin{equation}
  ds^2=\frac{1}{z^2}(dt^2-dz^2-dx^2-dy^2)
\end{equation}
by the recipe\cite{Kawamoto2022}
\begin{equation}
  \begin{split}
  &\sqrt{r^2+a^2}\cos\bar{t}=\frac{1}{2z}(a^2+x^2+y^2+z^2-t^2)
  \\
  &\sqrt{r^2+a^2}\sin\bar{t}=\frac{at}{z}
  \\
  &r\sin\theta\cos\phi=\frac{ax}{z}
  \\
  &-r\sin\theta\sin\phi=\frac{ay}{z}
  \\
  &-r\cos\theta=\frac{1}{2z}(-a^2+x^2+y^2+z^2-t^2),
  \end{split}
\end{equation}
where $a^2=\frac{3}{|\Lambda|}$ and the $\Lambda$ is absorbed. Alternatively, the AdS metric (1) can be obtained as a result of reflecting null-shells from a boundary \cite{Halilsoy2024}.
we can see that for constant z the AdS spacetime can be reduced to a $2+1-D$ flat spacetime, that is, every slice of constant z corresponds to a lower/flat spacetime.

Consider the Dirac equation in AdS spacetime by using the Newman-Penrose (NP) formalism \cite{Newman1962} refering to Chandrasekhar \cite{Chandrasekhar1983}.

\begin{equation}
  \begin{split}
  &(D+\epsilon-\rho)F_1+(\bar{\delta}+\pi-\alpha)F_2=i mG_1
  \\
  &(\Delta+\mu-\gamma)F_2+(\delta+\beta-\tau)F_1=i mG_2
  \\
  &(D+\bar{\epsilon}-\bar{\rho})G_2-(\delta+\bar{\pi}-\bar{\alpha})G_1=i mF_2
  \\
  &(\Delta+\bar{\mu}-\bar{\gamma})G_1-(\bar{\delta}+\bar{\beta}-\bar{\tau})G_2=i mF_1,
  \end{split}
\end{equation}
where $\alpha, \beta, \gamma, \rho, \epsilon, \mu, \pi, \tau $ are spin coefficients, $D, \Delta, \delta$, the directional derivatives, m the mass and $F_1, F_2, G_1, G_2$ the spinor components. From the Eqn (1), we may choose the null tetrad basis 1-forms as
\begin{equation}
  \begin{split}
  &\sqrt{2}l=\frac{1}{z}(dt-dz)
  \\
  &\sqrt{2}n=\frac{1}{z}(dt+dz)
  \\
  &\sqrt{2}m=\frac{1}{z}(dx+idy)
  \\
  &\sqrt{2}\bar{m}=\frac{1}{z}(dx-idy),
  \end{split}
\end{equation}
which gives us the only non-zero spin coefficients $\epsilon=\gamma=-\frac{1}{2\sqrt{2}}, \mu=\rho=\frac{1}{\sqrt{2}}$ ,Ricci scalar $R=-12$ and the directional derivatives
\begin{equation}
  \begin{split}
  &D=\frac{z}{\sqrt{2}}(\partial_t+\partial_z)
  \\
  &\Delta=\frac{z}{\sqrt{2}}(\partial_t-\partial_z)
  \\
  &\delta=-\frac{z}{\sqrt{2}}(\partial_x+i\partial_y)
  \\
  &\bar{\delta}=-\frac{z}{\sqrt{2}}(\partial_x-i\partial_y).
  \end{split}
\end{equation}
Therefore, we have the Dirac equations
\begin{equation}
  \begin{split}
  &(D-\frac{3}{2\sqrt{2}})F_1+\bar{\delta}F_2=i mG_1
  \\
  &(\Delta+\frac{3}{2\sqrt{2}})F_2+\delta F_1=i mG_2
  \\
  &(D-\frac{3}{2\sqrt{2}})G_2-\delta G_1=i mF_2
  \\
  &(\Delta+\frac{3}{2\sqrt{2}})G_1-\bar{\delta}G_2=i mF_1.
  \end{split}
\end{equation}

To solve these equations, we consider the ansatz
\begin{equation}
   \begin{split}
        & F_i=f_i(z)e^{i\omega t+ip_1 x+ip_2 y}
        \\
        & G_i=g_i(z)e^{i\omega t+ip_1 x+ip_2 y}.
   \end{split}
\end{equation}
Therefore, we derive
\begin{equation}
  \begin{split}
       &  Pf_1-(p_2+ip_1)f_2=\frac{i\sqrt{2}mg_1}{z}
       \\
       &  \bar{P}f_2-(p_2-ip_1)f_1=-\frac{i\sqrt{2}mg_2}{z}
       \\
       &  Pg_2-(p_2-ip_1)g_1=\frac{i\sqrt{2}mf_2}{z}
       \\
       &  \bar{P}g_1-(p_2+ip_1)g_2=-\frac{i\sqrt{2}mf_1}{z},
  \end{split}
\end{equation}
with the definition of operator, $P=\partial_z-\frac{3}{2z}+i\omega$ and its complex conjugate $\bar{P}$. By multiplying all the equations with its complex conjugate operator, we obtain
\begin{equation}
  \begin{split}
       &  f''_1-\frac{3}{z}f'_1+(\frac{15-8m^2}{4z^2}+\omega^2-p^2_1-p^2_2)f_1=-\frac{i\sqrt{2}m}{z^2}g_1
       \\
       &  f''_2-\frac{3}{z}f'_2+(\frac{15-8m^2}{4z^2}+\omega^2-p^2_1-p^2_2)f_2=\frac{i\sqrt{2}m}{z^2}g_2
       \\
       &  g''_2-\frac{3}{z}g'_2+(\frac{15-8m^2}{4z^2}+\omega^2-p^2_1-p^2_2)g_2=-\frac{i\sqrt{2}m}{z^2}f_2
       \\
       &  g''_1-\frac{3}{z}g'_1+(\frac{15-8m^2}{4z^2}+\omega^2-p^2_1-p^2_2)g_1=\frac{i\sqrt{2}m}{z^2}f_1,
  \end{split}
\end{equation}

Now consider the on-shell condition  and set $\omega^2-p^2_1-p^2_2=p^2_3+m^2=k^2$. One observes that from the equations the choices $f_1=ig_1$ and $f_2=ig_2$ decouple the equations and the Dirac equations are in consistent with each other. To solve the Dirac equations, we consider
\begin{equation}
       \begin{split}
       & f_1=A+iB \\
       & f_2=C+iD \\
       & g_1=E+iF \\
       & g_2=G+iH \\
       \end{split}
\end{equation}
where A, B,... are real functions and set $\Box f=f''-\frac{3}{z}f'+(\frac{15-8m^2}{4z^2}+k^2)f$. We then derive
\begin{equation}
       \begin{split}
       & \Box A=\frac{\sqrt{2}m}{z^2}F,  \Box B=-\frac{\sqrt{2}m}{z^2}E \\
       & \Box C=-\frac{\sqrt{2}m}{z^2}H,  \Box D=\frac{\sqrt{2}m}{z^2}G \\
       & \Box G=\frac{\sqrt{2}m}{z^2}D,  \Box H=-\frac{\sqrt{2}m}{z^2}C \\
       & \Box E=-\frac{\sqrt{2}m}{z^2}B,  \Box F=\frac{\sqrt{2}m}{z^2}A. \\
       \end{split}
\end{equation}
Equation (11) plays the crucial to get the consistent Dirac equaitons. However, to decouple the equations, one may choose
\begin{equation}
       \begin{split}
       & A=-F, C=-H \\
       & B=E, D=G. \\
       \end{split}
\end{equation}
Thus,
\begin{equation}
       \begin{split}
       & f_1=A+iB, f_2=C+iD \\
       & g_1=B-iA, g_2=D-iC, \\
       \end{split}
\end{equation}
and $f_1=ig_1$, $f_2=ig_2$ are satisfied. The equation (11) should be reduced to $\Box A=\frac{\sqrt{2}m}{z^2}A$, $\Box B=-\frac{\sqrt{2}m}{z^2}B$ or
\begin{equation}
       \begin{split}
       & A''-\frac{3}{z}A'+(\frac{15-8m^2-4\sqrt{2}m}{4z^2}+k^2)A=0 \\
       & B''-\frac{3}{z}B'+(\frac{15-8m^2+4\sqrt{2}m}{4z^2}+k^2)B=0. \\
       \end{split}
\end{equation}

To solve A and B, we refer to Lommel \cite{Lommel1871} that
\begin{equation}
  z^2f''+(2\alpha-2\beta\nu+1)zf'+[\alpha(\alpha-2\beta\nu)+\beta^2\gamma^2z^{2\beta}]f=0
\end{equation}
with a solution
\begin{equation}
 f=z^{\beta\nu-\alpha}[a_0 J_\nu(\gamma z^\beta)+b_0 J_{-\nu}(\gamma z^\beta)].
\end{equation}
Therefore,
\begin{equation}
  A(z)=z^2[a_0 J_{\nu_1}(kz)+b_0 J_{\nu_2}(kz)],
\end{equation}
where $\nu_{1,2}=\pm\sqrt{\frac{1}{4}+\sqrt{2}m(\sqrt{2}m+1)}$, and
\begin{equation}
  B(z)=z^2[c_0 J_{\nu_1}(kz)+d_0 J_{\nu_2}(kz)]
\end{equation}
with $\nu_{1,2}=\pm\sqrt{\frac{1}{4}+\sqrt{2}m(\sqrt{2}m-1)}$ and ($a_0, b_0, c_0, d_0$) constants. For massless case, $\nu=\pm\frac{1}{2}$, we got $z^2 J_{\pm\frac{1}{2}}$. The Bessel function performs as a periodic function with decay. We may consider that asymptotic approach, $z>>1$, the differential equation reduces to
\begin{equation}
  f''-\frac{3}{z}f'+k^2f=0,
\end{equation}
which gives a neat solution, $z^2J_2(kz)$ or $z^2N_2(kz)$, where $N(kz)$ is the Neumann function. Therefore, from our ansatz (7), we can build the spinor.

One may also consider the near-field effect, the differential equation reduces to
\begin{equation}
  f''+(\frac{15-8m^2-4\sqrt{2}m}{4z^2}+k^2)f=0.
\end{equation}
Again from Lommel, we see
 $f(z)=z^\frac{1}{2}[a_0 J_{\nu_1}(kz)+b_0 J_{\nu_2}(kz)]$,
with $\nu_{1,2}=\pm\sqrt{\frac{-7}{2}+\sqrt{2}m(\sqrt{2}m+1)}$, and ($a_0, b_0$) constants. We know the Bessel functions would be approximate to $\frac{1}{\sqrt{z}}cos(kz+phase)$ and $\frac{1}{\sqrt{z}}sin(kz+phase)$. $\therefore f\approx a_0sin(kz)+b_0cos(kz)+const.$, a plane wave. 
If we regard these solutions as neutrinos. For simplicity, we let 2 mass states to be
\begin{equation}
   \begin{split}
        &f_{\nu 1}=A_0sin(k_1z)\\
        &f_{\nu 2}=A_0sin(k_2z),
   \end{split}
\end{equation}
where $k_1=\sqrt{p^2_3+m^2_1}\approx p_3+\frac{m^2_1}{2p_3}$ and $k_2\approx p_3+\frac{m^2_2}{2p_3}$. Set $k=\frac{k_1+k_2}{2}$ and $\Delta k=\frac{m^2_1-m^2_2}{4p_3}$, then $k_1=k+\Delta k$ and $k_2=k-\Delta k$. The combination of neutrinos becomes
\begin{equation}
  \nu_1+\nu_2=2A_0sin(kz)cos(\Delta kz).
\end{equation}
The neutrino beat or neutrino oscillation is possible to be seen in the AdS background.

The AdS spacetime is an attractive geometry as it is well-known from AdS/CFT correspondence \cite{Maldacena1999}. Particles tend to deviate inward.
We can observe that $z^2$ accompanies all the Bessel functions, which makes all the wavefunctions become zero as $z \to$ 0. There emerges a boundary at $z=0$. But $z=0$ is undefined in the metric (1). All the information would be lost in the boundary, the wavefunction would disappear. However, after examining the geodesic (in the Appendix), we find that the wavefunction can only diminish to a $z=z_0$ plane. We may choose the $\xi$ to be very large, so that $z_0=\frac{1}{\xi}$ would be in the edge of boundary $z=0$ (in the Appendix). Therefore, when $z \to z_0$, $z^2_0J(kz_0)$ becomes constant. The other components still propagate in the $z_0$ plane, the $2+1-D$ lower and flat spacetime, see Figure 1.


\section{Tachyonic condition}

In this section we assume the tachyonic condition, that is,
\begin{equation}
  \omega^2-p^2_1-p^2_2=p^2_3+m^2=-\kappa^2.
\end{equation}
Obviously, $p^2_3>0$. So $m$ has to be replaced by $im$ for $\kappa^2 >0$. The equations of motion now become
\begin{equation}
  \begin{split}
       & \bar{\Box} f_1=\frac{\sqrt{2}m}{z^2}g_1,  \bar{\Box} f_2=-\frac{\sqrt{2}m}{z^2}g_2 \\
       & \bar{\Box} g_1=-\frac{\sqrt{2}m}{z^2}f_1,  \bar{\Box} g_2=\frac{\sqrt{2}m}{z^2}f_2 \\,
  \end{split}
\end{equation}
where $\bar{\Box}f=f''-\frac{3}{z}f'+(\frac{15+8m^2}{4z^2}-\kappa^2)f$. The condition $f_1=ig_1$, $f_2=ig_2$ also applies here, but the function can not be decomposed into real and imaginary part easily. We have to directly solve equations
\begin{equation}
       \bar{\Box} f_1=-i\frac{\sqrt{2}m}{z^2}f_1,  \bar{\Box} f_2=i\frac{\sqrt{2}m}{z^2}f_2,
\end{equation}
The solutions would be $z^2 K_{\nu_{1,2}}(\kappa z)$, where $K_\nu(\kappa z)$ is modified Bessel function, and $\nu_{1,2}=\pm\frac{1}{2}\mp i\sqrt{2}m$ for $f_1$, and $\nu_{1,2}={\pm\frac{1}{2}\pm i\sqrt{2}m}$ for $f_2$.

If the neutrinos are tachyons \cite{Chodos2023}, which means the mass is negligible. The modified Bessel function would become $K_{\pm\frac{1}{2}}(\kappa z)$. The wavefunction $z^2 K_{\nu}(\kappa z)$ then be shown in Figure 2. Before the wavefunction diminishes to the boundary, it has a maximal chance to be seen at constant slice $z_d$. The particles here are fermions, therefore they are more likely to clump in a $z=z_d$ slice, that is, a $2+1- D$ spacetime. There arises a phenomenon: In AdS spacetime, we have found a special shell that tachyonic neutrinos inhabit. In other words, we find a "Flatland" for neutrinos, since the shell is a $2+1-D$ flat spacetime. It is important to find a flat frame embedded in the curved spacetime. Thus, in curved AdS background, there may exist a flat shell, and in this shell, one could discuss the propagation with no curvature effects.




\begin{figure}
  \centering
  \includegraphics[width=0.3\textwidth]{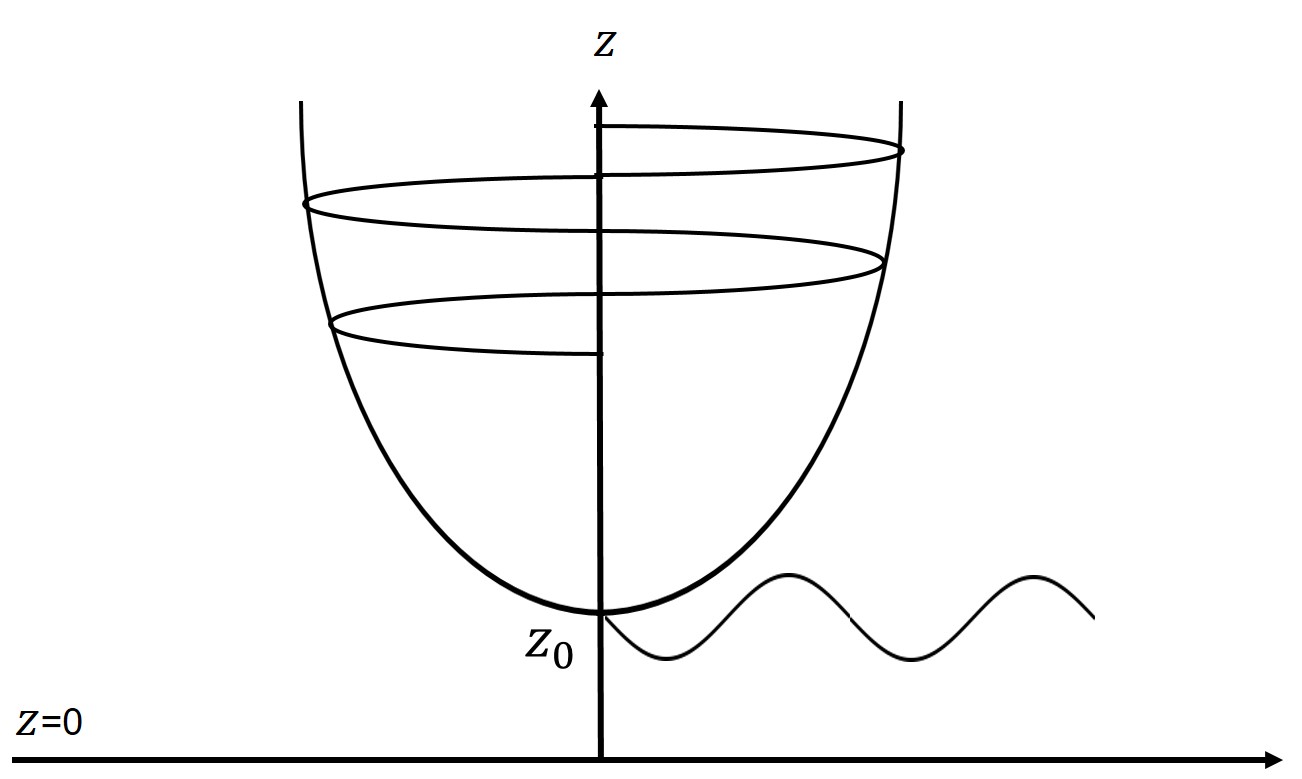}
  \caption{The $3+1-D$ wavefunction diminishes and collapses into a $2+1-D$ plane wave at $z=z_0$.}\label{f2}
\end{figure}

\begin{figure}
  \centering
  \includegraphics[width=0.3\textwidth]{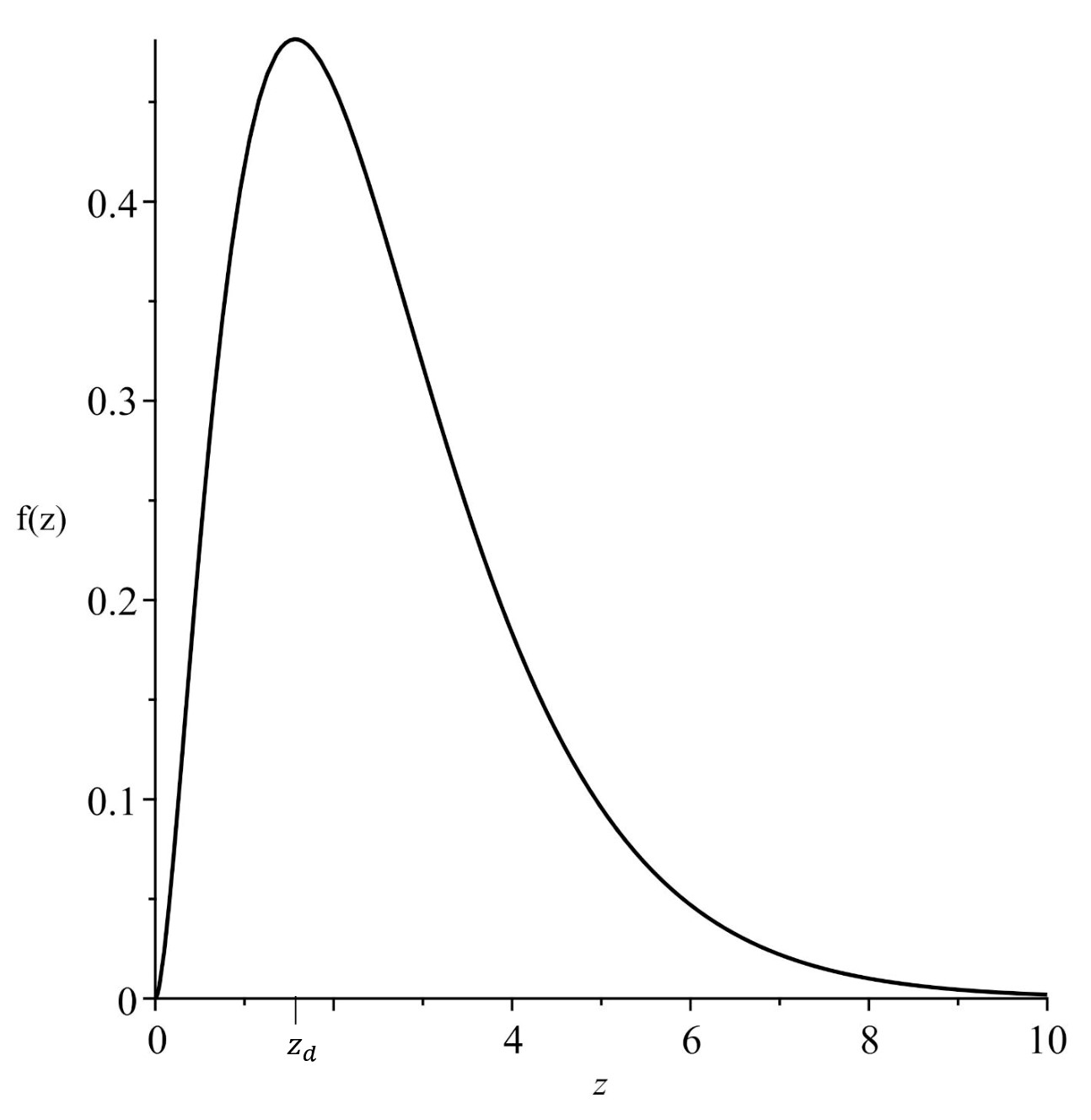}
  \caption{The wavefunction of tachyon has a maximum at $z=z_d$.}\label{f2}
\end{figure}

\section{Majorana condition}

In Dirac equations, we may impose the Majorana condition \cite{Aitchison2004},
\begin{equation}
  \Psi_M=\begin{bmatrix}
           \phi \\ i\sigma_2\phi^*
         \end{bmatrix},
\end{equation}
such that $\Psi_M$ would be its own antiparticle, where $\sigma_2=\begin{bmatrix}
           0 & -i\\ i & 0
         \end{bmatrix}$. Set $\phi=\begin{bmatrix}
           A \\ B
         \end{bmatrix}$.
\begin{equation}
\therefore \Psi_M=\begin{bmatrix}
           A\\B\\B^*\\-A^*
         \end{bmatrix}.
\end{equation}
Comparing Chandrasekhar's assumption \cite{Chandrasekhar1983} with the Majorana condition leads to
\begin{equation}
         \begin{bmatrix}
           P^A\\ \bar{Q}^{A'}
         \end{bmatrix}
         =
         \begin{bmatrix}
           P^0 \\ P^1 \\ \bar{Q}^{0'}\\ \bar{Q}^{1'}
         \end{bmatrix}
         =
         \begin{bmatrix}
           F_1 \\ F_2 \\ -G_2\\ G_1
         \end{bmatrix}
         =
         \begin{bmatrix}
           A \\ B \\ B^*\\ -A^*.
         \end{bmatrix}
\end{equation}
In the equation, $A\rightarrow F_1, B\rightarrow F_2, B^*\rightarrow -G_2, -A^*\rightarrow G_1$, we therefore find the Majorana condition for Dirac equation in AdS spacetime,
\begin{equation}
  F_1=-G_1^*, F_2=-G_2^*,
\end{equation}
and the condition of self complex conjugate,
\begin{equation}
  F_i=F_i^*, G_i=G_i^*
\end{equation}
We cast the ansatz as
\begin{equation}
   \begin{split}
        & F_i=f_i(z)e^{-\omega t-p_1 x-p_2 y}
        \\
        & G_i=g_i(z)e^{-\omega t-p_1 x-p_2 y}.
   \end{split}
\end{equation}
Bringing (31) into (3) gives
\begin{equation}
   \begin{split}
        &  Mf_1+(p_1-ip_2)f_2=\frac{i\sqrt{2}mg_1}{z}
       \\
       &  \bar{M}f_2+(p_1+ip_2)f_1=\frac{i\sqrt{2}mg_2}{z}
       \\
       &  Mg_2+(-p_1-ip_2)g_1=\frac{i\sqrt{2}mf_2}{z}
       \\
       &  \bar{M}g_1+(-p_1+ip_2)g_2=\frac{i\sqrt{2}mf_1}{z},
   \end{split}
\end{equation}
where we define $M=-\omega+\partial_z-\frac{3}{2z}$ and $\bar{M}=-\omega-\partial_z-\frac{3}{2z}$, and multiply $M$ and $\bar{M}$ for each  as we did previously. The differential equations read
\begin{equation}
   \begin{split}
        &  f''_1-\frac{3\omega}{z}f_1-(\frac{3+8m^2}{4z^2}+k^2)f_1=\frac{i\sqrt{2}m}{z^2}g_1
       \\
       &  f''_2-\frac{3\omega}{z}f_2-(\frac{15+8m^2}{4z^2}+k^2)f_2=\frac{i\sqrt{2}m}{z^2}g_2
       \\
       &  g''_1-\frac{3\omega}{z}g_1-(\frac{15+8m^2}{4z^2}+k^2)g_1=\frac{i\sqrt{2}m}{z^2}f_1
       \\
       &  g''_2-\frac{3\omega}{z}g_2-(\frac{3+8m^2}{4z^2}+k^2)g_2=\frac{i\sqrt{2}m}{z^2}f_2,
   \end{split}
\end{equation}
and also the on-shell condition applies here. The difference between $f_i$ and $g_i$ are due to that $M$ and $\bar{M}$ do not commute.
Then considering (29) and (30), $g_1\rightarrow -f_1^*\rightarrow -f_1$, we obtain the Whittaker equation (see \cite{Arfken2013})

\begin{equation}
   \begin{split}
        f''_1-\frac{3\omega}{z}f_1-(\frac{3+8m^2-i4\sqrt{2}m}{4z^2}+k^2)f_1=0
   \end{split},
\end{equation}
and similar for the rest equations.

We shall discuss some special cases in these equations. 
For heavily sterile Majorana particles in seesaw mechanism \cite{Bilenky2018}, $m>>1$ and $z<<1$, eqn (34) can be reduced to
\begin{equation}
   \begin{split}
        f''_1-\frac{2m^2}{z^2}f_1\approx 0.
   \end{split}
\end{equation}
The wavefunction of Majorana particles is
\begin{equation}
        f_1=a_0 e^{k_1\ln z}+b_0 e^{k_2\ln z},
\end{equation}
where $k_{1,2}=\frac{1\pm\sqrt{1+8m^2}}{2}$ with $a_0 = b_0 = const.$, and similarly for the $g_i$ functions but picking up a minus sign. Then $f_i=-g_i^*$ and $f_i=f_i^*$ fit the criteria (29) and (30).

We shall make another approximation. 
One can also consider that $f_1$ is a wavefunction consisted of a pair of fermion and anti-fermion in trapped ion experiments with a reasonable assumption, $\omega <<1$. In addition, as we restore the Planck constant $\hbar$ and speed of light c, one can see that $\frac{c}{\hbar}$ is a big number. Therefore, we may approximately regard the Whittaker equations of equation (34) as

\begin{equation}
   \begin{split}
        f''_1-(\frac{\frac{8m^2 c^2}{\hbar^2}-i4\sqrt{2}\frac{mc}{\hbar}}{4z^2}+k^2)f_1=0
   \end{split},
\end{equation}
and the Majorana tachyon, $m \to im$ and $k^2 \to -\kappa^2$,
\begin{equation}
   \begin{split}
        f''_1-(\frac{-\frac{8m^2 c^2}{\hbar^2}+4\sqrt{2}\frac{mc}{\hbar}}{4z^2}-\kappa^2)f_1=0
   \end{split}.
\end{equation}
We would like to solve eqn (38) first. Again, with the help of Lommel's method\cite{Lommel1871}, the solution is $f_1(z)=z^\frac{1}{2}[a_0 J_{\nu_1}(\kappa z)+b_0 J_{\nu_2}(\kappa z)]$ with $\nu_{1,2}=\pm\frac{1}{2}\sqrt{1-4\sqrt{2}\bar{m}(\sqrt{2}\bar{m}-1)}$, termed $\bar{m}=\frac{mc}{\hbar}$ and ($a_0, b_0$) constants. One may also see that $f_i=-g_i^*$ and $f_i=f_i^*$ in (29) and (30). This Majorana tachyon in ion trap seems moving smoothly as plane waves.

To Solve eqn (37), we take $ z \to iz$, then Bessel turns into modified Bessel. The solution would be
\begin{equation}
   \begin{split}
       f_1(z)=z^\frac{1}{2}[a_0 K_{\nu_1}(kz)+b_0 K_{\nu_2}(kz)],
   \end{split}
\end{equation}
with $\nu_{1,2}=\pm \frac{1}{2} (A\mp iB$), ($a_0, b_0$) constants, $A=(\frac{1+8\bar{m}^2+\sqrt{(1+8\bar{m}^2)^2+32\bar{m}^2}}{2})^\frac{1}{2}$ and $B=(\frac{-(1+8\bar{m}^2)+\sqrt{(1+8\bar{m}^2)^2+32\bar{m}^2}}{2})^\frac{1}{2}$. To fit the self-complex conjugation, one should regard the field as a nearly massless particle in the modified Bessel function. That is, the Majorana condition naturally restricts the particle's mass. However, it is this very tiny mass that triggers the hopping.
The Majorana particles of equation (39) act like Dirac tachyons. They form flatlands as Figure 2 showed. The difference is that we may regard them as Majorana neutrinos. Since the Majorana condition is quasi-satisfied with the nearly zero mass and the neutrinos have truly tiny masses. With this argument, there raises a phenomenon. We see that the parameter A of index $\nu_{1, 2}$ in modified Bessel includes mass and then $z^\frac{1}{2} K_{\nu}(kz)$ is a function of mass, although $m<<1$. Therefore as the mass varies, the $z_d$ shifts, see Figure 2. We may expect that in the AdS spacetime the Majorana neutrinos hop between different $z_d$ since they vary masses in traveling. Therefore, neutrinos jump around different flatlands. This opens a possibility for the quest: Can we mimic the D-brane stack and open strings \cite{Dai1989} in the lab? Since neutrinos here play the role of strings to connect every flatland.

One may notice a concept. Spacetime like AdS can be projected to the Ginzburg-Landau theory \cite{Wu2016}. Vice versa, quantum mechanical materials can simulate spacetime \cite{Viermann2022}. Additionally, the ion trap can imitate the Dirac fermions \cite{Gerritsma2010} and Dirac tachyons \cite{Lee2015}. In these ideas, vector potential plays the central role. There is a way to connect them. We may consider the electromagnetic interaction and gravitational effect in the derivative of the spinor $\Psi$\cite{Atiyah1984, Mukhopadhyay2000}, $D_\mu\Psi=\partial_\mu\Psi+iq_1 A_\mu \Psi+q_2\Gamma_\mu \Psi$, where $A_\mu$ and $\Gamma_{\mu}$ are vector potential and Riemannian connection with coupling constant $q_1$ and $q_2$, respectively. We could find a proper $A_\mu$ to replace $\Gamma_{\mu}$, or we can say, we mimic the non-flat space by $A_\mu$. Therefore, geometric constraints have been replaced by  potential constraints. Since the trapped ions can vary the paricles’s mass and are controllable, to test the D-brane stack and open strings in this way might be possible.



\section{Discussion and Conclusion}

By exploiting the conformally flat spacetime, we find a general form of Dirac fermions in AdS spacetime, which can be applied to free neutrinos, electrons, quarks and especially the spin-$\frac{1}{2}$ ions. A suitable $A_\mu$ may simulate the AdS spacetime. Therefore an emergence of effective Dirac tachyons in the lab \cite{Lee2015} may be imposed to test the flatland in a 2+1-D case. Since the trapped ions are feasible to mimic fermions, we may go further to investigate Majorana particles in the lab. And in our study, rectangular + time coordinate with one variable, z, makes the design of ion trap easier and feasible. Finally, the Whittaker equations of equation (34) deserve further attentions to investigate other interesting phenomena in this context.




\section*{Appendix}

Since we have
\begin{equation}
  ds^2=\frac{1}{z^2}(dt^2-dz^2-dx^2-dy^2),
\end{equation}
the geodesic lagrangian is
\begin{equation}
  L=\frac{1}{2z^2}(\dot{t}^2-\dot{z}^2-\dot{x}^2-\dot{y}^2).
\end{equation}
Because x, y, and t parts are cyclic, we have
\begin{equation}
  \begin{split}
       & \frac{\partial L}{\partial \dot{t}}=\frac{\dot{t}}{z^2}=\alpha_0 \\
       & \frac{\partial L}{\partial \dot{x}}=\frac{\dot{x}}{z^2}=\beta_0 \\
       & \frac{\partial L}{\partial \dot{y}}=\frac{\dot{y}}{z^2}=\gamma_0,
  \end{split}
\end{equation}
where a 'dot' denotes derivative with respect to the proper time $\tau$.
As for the z part, we should consider $1=\frac{1}{z^2}(\alpha_0^2z^4-\dot{z}^2-\beta_0^2z^4-\gamma_0^2z^4)$, therefore
\begin{equation}
  \dot{z}^2=-z^2+(\alpha_0^2-\beta_0^2-\gamma_0^2)z^4=\xi^2z^4-z^2,
\end{equation}
where $\xi^2=\alpha_0^2-\beta_0^2-\gamma_0^2$.
Also, $\frac{dz}{d\tau}=\frac{dz}{dt}\frac{dt}{d\tau}=\frac{dz}{dt}\alpha_0z^2$. As a result,
$\frac{dz}{dt}=\frac{1}{\alpha_0}\sqrt{\frac{\xi^2z^2-1}{z^2}}$, and brings us
\begin{equation}
  z=\sqrt{\frac{(t-t_0)^2\xi^2}{\alpha_0^2}+\frac{1}{\xi^2}},
\end{equation}
with constant $t_0$. We see that the minimum of z, $z_0=\frac{1}{\xi}$.

Data Availability Statement: No Data associated in the manuscript.




\end{document}